\documentclass[aps,prd,reprint,a4paper,showpacs,nofootinbib,superscriptaddress]{revtex4-2}

\usepackage{hyperref}
\hypersetup{
  colorlinks=true,        
  linkcolor=blue,         
  citecolor=cyan,         
}

\usepackage{graphicx}
\usepackage{dcolumn}
\usepackage{bm}
\usepackage{color}
\usepackage{enumitem}
\usepackage{amsmath}

\usepackage{amssymb}

\newcommand{\beq}{\begin{equation}}
\newcommand{\eeq}{\end{equation}}
\newcommand{\bea}{\begin{eqnarray}}
\newcommand{\eea}{\end{eqnarray}}

\usepackage{bbm}
\usepackage{amsfonts}
\usepackage{mathrsfs}
\usepackage{latexsym}
\usepackage{epsfig}
\usepackage{epstopdf}
\usepackage{epstopdf}
\usepackage{graphicx}
\usepackage{amssymb}
\usepackage{amsmath}
\usepackage{dcolumn}
\usepackage{bm}
\usepackage{comment}
\usepackage{xcolor}
\usepackage{dsfont}
\usepackage{amsmath}
\usepackage{booktabs}
\usepackage{float, geometry,lipsum}
\usepackage[T1]{fontenc}
\usepackage[utf8]{inputenc}   
\usepackage{lmodern}
\usepackage{cmap}
\begin{document}

\title{Topological analysis of novel charged-AdS black hole solutions embedded in dark matter halos with quintessence field}

\author{Muhammad Yasir}
\email{yasirciitsahiwal@gmail.com; muhammadyasir@nbu.edu.cn}
\affiliation{Institute of Fundamental Physics and Quantum Technology, Department of Physics, School of Physical Science and Technology, Ningbo University, Ningbo, 315211, Zhejiang, People’s Republic of China}

\author{Hafiza Bushra Mujahid}
\email{bushrashahzad272@gmail.com}
\affiliation{Department of Mathematics, COMSATS University Islamabad, Sahiwal Campus-57000, Islamabad, Pakistan}

\author{Faisal Javed}
\email{Faisal@tnu.edu.cn}
\affiliation{Department of Physics, Tongji University, Shanghai 201804, China}

\author{Shahid Qaisar}
\email{shahidqaisar90@ciitsahiwal.edu.pk}
\affiliation{Department of Mathematics, COMSATS University Islamabad, Sahiwal Campus-57000, Islamabad, Pakistan}

\begin{abstract}
In this work, we investigate the properties of a charged anti-de Sitter (AdS) black hole in the presence of a NGC 4649-type dark matter halo and a surrounding quintessence field. Starting from the modified Einstein field equations (EFEs), we construct the corresponding black hole solutions by incorporating the effects of electric charge, cosmological constant, dark matter distribution, and quintessence characterized by a variable equation of state. We analyze the resulting spacetime geometry through the metric function and horizon structure, showing that the dark matter halo strengthens the gravitational attraction, whereas the quintessence field provides a repulsive contribution that substantially alters the asymptotic behavior of the spacetime. Furthermore, the thermodynamic properties are investigated by applying the Duan-\(\phi\)-mapping topological theory together with the generalized free energy method. This framework allows us to examine the topological structure, critical behavior, and stability conditions of the black hole system in a consistent manner.
\textcolor{black}{
We further study the fractional shifts relative to the reference value of
\(\omega_q\) and observe that variations in \(\omega_q\) produce systematic changes which highlight the important role of the quintessence field in determining the phase structure and critical behavior of the black hole system.}
\\\\
Keywords; Charged-AdS black hole models, Quintessence contribution, Dark matter halos, Duan's $(\phi)$-mapping theory, Thermodynamic ensembles
\end{abstract}
\maketitle
\date{\today}

\section{Introduction}
\label{introduction}

Observational evidence indicates that nearly every massive galaxy harbors a supermassive black hole at its center \cite{G1}. These compact objects reside in complex astrophysical environments, typically surrounded by dark matter halos and, in many scenarios, influenced by the large-scale dynamics of the Universe. Consequently, the motion of matter, radiative processes, and a wide range of optical phenomena have been extensively investigated in the framework of black holes embedded in dark matter halos \cite{G2,G3,G4,G5,G6}. 

Over the past decades, a wealth of astrophysical and cosmological observations has established that the Universe is predominantly composed of dark matter and dark energy. Measurements of the cosmic microwave background indicate that dark matter and dark energy account for approximately 26.8\% and 68.3\% of the total energy density of the Universe, respectively, whereas ordinary baryonic matter contributes only about 4.9\% \cite{GR7}. The presence of dark matter plays a fundamental role in astrophysics and cosmology, providing a consistent explanation for a broad range of phenomena, including the formation of large-scale cosmic structures \cite{GR8}, the dynamics of bullet clusters \cite{GR9}, galactic rotation curves \cite{GR10}, and gravitational lensing, all of which are difficult to reconcile within the framework of visible matter alone. Moreover, recent observations from X-ray and cosmic-ray missions, including Chandra, Fermi-LAT, AMS, and DAMPE, have reported anomalous excesses in high-energy photons and charged particles \cite{GR11}. The enhanced dark matter density in the vicinity of galactic centers has been widely proposed as a plausible mechanism for explaining these observational anomalies.

Recent astronomical observations provide compelling evidence for the accelerating expansion of the Universe, indicating the presence of a state with negative pressure.  The negative pressure responsible for cosmic acceleration may originate from two distinct sources: the cosmological constant or a dynamical dark energy component, commonly referred to as quintessence. The latter is characterized by the equation of state
$p_q = w_q \rho_q$, where the equation-of-state parameter satisfies
$-1 < w_q < -\frac{1}{3}$, thereby generating the negative pressure required to drive the accelerated expansion of the Universe. In the limiting case \(w_q=-1\), the quintessence model reduces to the cosmological constant.
As recognized in previous studies \cite{QI1,QI2,QI3}, the acceleration poses a challenge for consistent quantum gravity theory, particularly because of the outer horizon. This horizon prevents the introduction of the observable S-matrix in terms of asymptotic past and future states. The outer horizon of de Sitter space is distinctly different from the inner horizon of a black hole, which is located far from the black hole in asymptotically flat space.

Quintessence acts as a smooth dark energy component with negative pressure, which slows the growth of dark matter perturbations by opposing gravitational collapse. As a result, the overall abundance of dark matter halos is reduced, and halos tend to form later compared to the standard cold dark matter scenario. Regarding the physical size of halos, the absence of coupling means that quintessence does not exert any additional force on dark matter particles. Instead, its effect is indirect, leading to halos that are generally less concentrated and more extended, since the weakened gravitational growth prevents matter from collapsing as efficiently. Therefore, in the uncoupled case, changes in halo size and structure are primarily driven by the background cosmological dynamics rather than local interactions between dark matter and the quintessence field.

A more realistic astrophysical description of such systems requires incorporating both the effects of extended dark matter distributions and the cosmological constant, which gives the spacetime an asymptotically de Sitter geometry.
Although astrophysical black holes are generally expected to carry negligible net electric charge, since any substantial charge is rapidly neutralized by the surrounding plasma through the preferential accretion of oppositely charged particles \cite{G7}, this picture can be modified in magnetized environments. In particular, a black hole embedded in a magnetized plasma may acquire a non-negligible induced electric charge through charge-separation mechanisms, exceeding the minimal equilibrium value predicted in vacuum \cite{G8}. Such an induced charge can significantly affect the dynamics of the surrounding plasma \cite{G9} and has been proposed as a possible explanation for the hot-spot flare activity observed in the vicinity of Sgr A* \cite{G10}.

\textcolor{black}{Motivated by these considerations, we investigate Schwarzschild black hole solutions embedded in a DM halo whose density profiles are constrained by observational findings for elliptical  Dehnen-(1,4,5/2) and 
galaxy NGC~4649 (M60) \cite{DN1,DN2,NGC1,NGC2,NGC3,NGC4,NGC5}, including data from globular cluster (GC) measurements, the Hubble Space Telescope (HST), and stellar kinematics. Based on this observationally constrained profile, we derive a distinct static spacetime solution characterized by the critical velocity \(V_c\), the black hole mass \(M\), and the halo core radius \(a\). The resulting geometry continuously reduces to the Schwarzschild spacetime~\cite{KS} in the limiting case \(V_c=0\) and \(a=0\).
Within the standard cosmological framework, DM plays a fundamental role in the formation and evolution of cosmic structures. Galaxies are widely believed to be embedded within extended DM halos, which largely determine their mass distribution and account for the observed galactic rotation curves. Consequently, understanding the nature, spatial distribution, and dynamical influence of DM remains one of the major challenges in modern astrophysics and cosmology \cite{G11,G12,G13,G14,G15,G16,G17}.}

Furthermore, the thermodynamical behavior of phase structure of black holes has been attracted the attention since the beginning of the thermodynamic of black holes. 
Bekenstein and Hawking proved that the black holes possess entropy and temperature \cite{Ref3}. The Hawking temperature is related to the surface gravity $\kappa$ as $T_H={\kappa}/{2\pi}$, where the surface gravity for a static black hole is defined by $\kappa=\frac{1}{2}f'(r_h).$ Thus, the Hawking temperature becomes $T_H={f'(r_h)}/{4\pi}$ within the standard Boltzmann-Gibbs framework, the entropy is governed by the Bekenstein-Hawking area law
\begin{equation}\label{S}
S=\frac{A}{4}=\pi r_h^{2},
\end{equation}
where $A$ is the area of the event horizon. In extended thermodynamics space, the critical behavior of AdS black hole have a great interest in recent years, where the thermodynamic pressure \(P\) of a black hole is associated with the cosmological constant \(\Lambda\) is defined as \cite{TC1,TC2,TC3,TC4,TC5,TC6,TC7,TC8}
\begin{equation}\label{3}
P=\frac{3}{8\pi l^{2}}=-\frac{\Lambda}{8\pi G},
\end{equation}
where \(G\) denotes the Newton's gravitational constant. In black hole thermodynamics, the pressure \(P\) is thermodynamically conjugate to the volume \(V\). Incorporating this relation into the first law of black hole thermodynamics \cite{Ref1,Ref2, TC9,TC10,TC11,TC12,MYAJ2}, one obtains the modified first law
\begin{equation}
dM=\sum_{i}Y_{i}dx_{i}+T\,dS+V\,dP,
\end{equation}
where \(M\) represents the black hole mass interpreted as enthalpy, \(S\) denotes the entropy, \(T\) is the Hawking temperature, and \(Y_i dx_i\) corresponds to additional chemical potential terms associated with the thermodynamic parameters of the system. To investigate black hole criticality, we adopt the framework of thermodynamic topology introduced in
\cite{TC2,TC3,TC6,TC7,TC8,TC9,TC10,TC11,TC12,TC13,TC14,TC15}, based on Duan’s \(\phi\)-mapping topological theory. This formalism provides a systematic method for locating and classifying thermodynamic critical points. Through this approach, black holes can be categorized into different topological classes exhibiting both standard and novel critical phenomena \cite{Ensml1,Ensml2,Ensml3,Ensml4,Ensml5}. The black hole temperature is expressed as a function of entropy \(S\), pressure \(P\), and other thermodynamic parameters \(x_i\):
\begin{equation}\label{T}
T = T(S, P, x_{i}).
\end{equation}
By eliminating the state parameter through the condition 
\((\partial_{S} T)_{P, x_{i}} = 0\), Duan’s thermodynamic potential 
\(\phi\) is introduced as \cite{TC6,TC7,TC16,TC17,TC18,TC19,TC20}
\begin{equation}\label{4}
\phi = (\sin\theta)^{-1} T(S, x_{i}).
\end{equation}
The factor \((\sin\theta)^{-1}\) naturally establishes the connection within the \(\phi\)-mapping construction. Correspondingly, the associated two-dimensional vector field  \(\Phi = (\Phi^{S}, \Phi^{\theta})\) is defined by \cite{Ensml6,Ensml7,Ensml8,Ensml9,Ensml10}
\begin{equation}
\Phi^{\theta} = \left(\frac{\partial \Phi}{\partial
\theta}\right)_{S, x_{i}}, \quad \Phi^{S} = \left(\frac{\partial
\Phi}{\partial S}\right)_{\theta, x_{i}}.
\end{equation}

This vector field possesses a fixed zero point located at 
\(\theta = \pi/2\). The corresponding topological charge \(W\) is evaluated through
\begin{equation}
W = \int j^{0} \, d^{2}x = \sum_{i=1}^{N} w_{i} = \sum_{i=1}^{N}
\beta_{i} n_{i},
\end{equation}
where \(j^{0}\) denotes the topological current density, \(w_i\) represents the winding number associated with the \(i\)-th zero of the vector field \(\Phi\), \(\beta_i\) is the Hopf index, and \(n_i\) corresponds to the Brouwer degree in Refs. \cite{TC17,TC18,TC21,TC22,TC23,TC24,TC25}. 

Besides Duan’s \(\phi\)-mapping formulation, another useful perspective examines black hole configurations in incomplete thermodynamic parameter spaces \cite{TC6,TC7,TC8,TC15,TC16,TC17}. In this approach, the deficiency of winding numbers is employed to analyze thermal stability, while the total winding number is interpreted as the topological charge of the system. Within this framework, the free energy \(F\) is written as
\begin{equation}\label{F0}
F = E \;-\;S \tau^{-1},
\end{equation}
where \(E\) denotes the potential energy, \(S\) is the entropy, and \(\tau\) is a time-like parameter. For fixed electric potential, the corresponding Legendre-transformed potential is commonly \(F = M - \frac{S}{\tau} - \Phi_e Q.\) The corresponding vector field is then constructed as
\begin{equation}\label{5}
\Phi = \left( \frac{\partial F}{\partial r_{+}}, -\cot\theta \,
\csc\theta \right),
\end{equation}
which also exhibits a zero point at \(\theta = \pi/2\). The normalized unit vector field is defined by
\begin{equation}
n^{a} = \frac{\Phi^{a}}{\|\Phi\|}, \quad \Phi^{1} = \Phi_{r}, \quad
\Phi^{2} = \Phi_{\theta},
\end{equation}
with \(a = 1,2\). In black hole thermodynamics, winding numbers arise naturally through topological analysis of thermodynamic quantities such as free energy, temperature, or horizon radius. These quantities are mapped into a topological space where the winding number characterizes the number of thermodynamic defects or distinct black hole phases. The sum of all winding numbers defines the total topological charge, which serves as a robust indicator of thermodynamic stability. Variations in the winding number correspond to changes in the phase structure and therefore signal phase transitions or modifications in thermal stability. Physically, a winding number equal to zero implies the absence of a stable phase, whereas a winding number of one indicates the existence of a single stable black hole phase. Consequently, the winding number provides an efficient topological criterion for classifying black hole phases and their stability properties \cite{TC16,TC17,TC18,TC19,TC20,TC21,TC22,TC23}. 

Operationally, winding numbers are evaluated at the zeros of the first component of the normalized vector field \(n^a\), namely at the zeros of \(n^1\) for a fixed control parameter \(\tau\). The sum of these individual contributions gives the total topological number, or equivalently the topological charge, of the system. In the standard formalism, the topological class is determined through \cite{TC6,TC7,TC8,TC15,TC16,TC17}
\begin{equation}\label{6}
W = (2 \pi )^{- 1} \Omega= \frac{1}{2\pi} \oint_{C} \epsilon_{ab} n^{a} \partial_{i} n^{b}
dx^{i}.
\end{equation}
Here, \(n^a\) represents the unit vector field derived from the thermodynamic potential, \(\epsilon_{ab}\) is the antisymmetric tensor, and the integration is performed over a closed contour \(C\) in thermodynamic parameter space. For practical evaluation, one introduces a family of closed curves parameterized by 
\(\vartheta \in (0,2\pi)\)
\begin{equation}\label{7}
\left\{
\begin{aligned}
r &= a\cos\vartheta + r_0,\\
\theta &= b\sin\vartheta + \frac{\pi}{2}.
\end{aligned}
\right.
\end{equation}

These relations describe elliptical contours centered at 
\((r_{0}, \pi/2)\) with semiaxes \((a,b)\). After specifying suitable values of \((a,b,r_0)\) to enclose the candidate topological defects, the thermodynamic field (or equivalently the relevant free-energy-derived quantities) is evaluated along the contour, and the corresponding line integral is computed. The sign of each contribution depends on the local rotational orientation of the vector field, where counterclockwise rotation contributes positively and clockwise rotation contributes negatively. The total topological charge \(W\), obtained by summing all enclosed contributions, classifies the phase structure of the black hole system. Any change in \(W\) indicates a topological transition associated with a thermodynamic phase transition. Finally, after selecting the parameters \((a,b,r_0)\), we compute \(F\) along the corresponding contours.

\textcolor{black}{ Motivated by the growing interest in understanding of black hole physics in realistic topological environments, where both dark matter and dark energy are expected to exert important influences, we examine the properties of a charged-AdS black hole  surrounded by a NGC 4649-type dark matter halo and a quintessence field. Although previous investigations have widely considered black holes within dark matter backgrounds and in the presence of quintessence, a detailed study that simultaneously includes these two dominant cosmic components together with electric charge and AdS geometry remains comparatively limited. In particular, it is important to investigate how the combined effects of the attractive behavior of dark matter and the repulsive influence of quintessence through Duan's $(\phi)$-mapping theory.
The primary aim of this work is to demonstrate the thermodynamic topology which gives a conventional analysis of black hole phase transition and the topological behavior of charged AdS black holes surrounded by dark matter halos in the presence of a quintessence field. The geometry and thermodynamic relations can be modified by these factors, and the thermodynamic ensemble (canonical and mixed canonical ensembles) are examined for the considered charged black hole and the effects of the halo model and quintessence parameters on the locations of thermodynamic defects and the winding numbers are determined. By employing Duan's $(\phi)$-mapping topological current theory and the framework of topological thermodynamic defects, we identify the zero points of the thermodynamic vector field, evaluate their winding numbers and topological charges, and classify the corresponding black hole phases.}

This study explores the thermodynamic topology of charged black hole models surrounded by dark matter halos in the presence of a quintessence field. Section \ref{introduction} introduces the physical motivation and significance of black holes influenced by dark matter and quintessence. Further, we discuss the contribution of the quintessence field and examine its fundamental analytical properties. Section \ref{02} presents the gravitational field equations for charged black holes embedded in dark matter halos under the influence of quintessence. In Section \ref{04} investigates the Dehnen-(1,4,5/2) halo density profile and its implications for black hole geometry, while the characteristics of the NGC 4649 dark matter halo model are analyzed in Section \ref{03}. In Section \ref{05}, we derive the corresponding Smarr relation and establish the thermodynamic framework of the system. The thermodynamic topology of the black hole in the canonical ensemble is examined in Section \ref{06}. Subsequently, Sections \ref{07}, \ref{08}, and \ref{09} present the topological thermodynamic defect structure of the black hole in the canonical and mixed canonical ensembles. Finally, Section \ref{010} summarizes the main results and discusses the physical implications and future prospects of the present study.

\section{Field Equations with dark matter halo in the presence of quintessence field}\label{02}

Our analysis begins with the most general static, spherically symmetric spacetime, characterized by the line element \cite{KS, BSV}
\begin{equation}\label{D1}
ds^{2}=- f(r)dt^{2}+
\frac{dr^{2}}{\Psi(r)}+r^{2} d\Omega^{2},
\end{equation}
where
\begin{equation} \nonumber
d\Omega^{2}=d\theta^{2}+\sin^{2}\theta d\varphi^{2}.
\end{equation}
Throughout this work, we adopt natural units in which (G=c=1)
\begin{equation} \label{g4}
f(r)=\Psi(r)= 1-\frac{2m(r)}{r}.
\end{equation}
We now incorporate the quintessence field, modelled as an anisotropic fluid with equation of state $p_q = \omega_q \rho_q$. Following the Kiselev construction \cite{Q1}, the stress-energy tensor components in the orthonormal frame are:
\begin{eqnarray}\nonumber
T^{(q)t}_t = T^{(q)r}_r = -\rho_q, \quad T^{(q)\theta}_\theta = T^{(q)\phi}_\phi = \frac{1}{2} \rho_q(3\omega_q+1).
\\
\end{eqnarray}
For a static, homogeneous quintessence fluid, conservation laws demand a power-law form for the energy density, $\rho_q = -\frac{3\omega_q c_q}{2} r^{-3(\omega_q+1)}$. Integrating the above equation then yields a contribution to the metric function of the form $-\frac{c_q}{r^{3\omega_q + 1}}$, where $c_q$ is an integration constant related to $\rho_0$.
\textcolor{black}{For $-\frac{1}{3}<\omega_q<0$, the quintessence contribution decays as $(r\rightarrow\infty)$ and therefore does not modify the leading asymptotic behavior determined by the cosmological constant. For $(\Lambda<0)$, the spacetime consequently remains asymptotically AdS, and the terminology ``charged AdS black hole'' is appropriate. On the other hand, for $-1<\omega_q<-\frac{1}{3}$, one has $(3\omega_q+1<0)$, and the quintessence contribution grows with increasing (r). Consequently, this term can modify the large-(r) behavior of the spacetime.} Therefore, we consider an electrically neutral environment, and the Maxwell equations take the form:
\begin{equation} \label{g5}
\frac{1}{\sqrt{-g}} \partial_\nu \left(\sqrt{-g} F^{\mu\nu} \right)=0, \quad F_{\mu\nu} \equiv \partial_\mu A_\nu-\partial_\nu A_\mu, 
\end{equation}
are reduced to the following form \cite{G20}. Thus, the electric field strength satisfies as: $V'(r)=-{Q}/{r^2}$, where $Q$ is the integration constant. We conclude that $Q$ is the electric charge. 
\textcolor{black}{The dynamics of spacetime are described by EFEs, which relate the geometric properties of spacetime to its matter-energy content through the stress-energy tensor. The modified form of the EFEs is given by \cite{KS, BSV,Q1,G20}
\begin{equation}\label{eq:EFE}
G_{\mu\nu} + \Lambda g_{\mu\nu}=\frac{8\pi G}{c^{4}}T^{(mod)}_{\mu\nu},
\end{equation}
where \(G^{\nu}_{\mu}\) is the Einstein tensor, and  $T^{(mod)}_{\mu\nu}$ is the modified electromagnetic energy-momentum tensor.}
Finally, we assume that the line element given in (\ref{D1}) must be the solution to the Einstein equations with anisotropic matter of density $\rho(r)$, only the tangential pressure $P(r)$, the Maxwell field given above, and cosmological constant $\Lambda$, respectively. Then, the nonzero components of the corresponding stress-energy tensor are
\begin{eqnarray} \label{g6}
T^{t}_{t}=-\rho(r)-\frac{Q^2}{r^4}-\rho_q,  \quad
T^{r}_{r}=P_r(r)-\frac{Q^2}{r^4}-\rho_q, 
\end{eqnarray}
\begin{small}
\begin{eqnarray}\nonumber
8\pi T^{\theta}_{\theta}=8\pi T^{\varphi}_{\varphi}=8\pi P_t(r)= 8\pi P(r)+\frac{Q^2}{r^4}+\frac{1}{2} \rho_q(3\omega_q+1).\\ 
\label{g8}
\end{eqnarray}
\end{small}
Within the framework of General Relativity, the determination of the metric functions requires specifying the matter source through the right-hand side of the EFEs. Nevertheless, we emphasize that for galactic dark matter halos, the underlying equation of state remains unknown. To circumvent this limitation, Ref.~\cite{MTND} employed an alternative reconstruction approach based on the observed tangential velocity profiles of stars in galaxies. The dynamics of spacetime are described by Einstein's field equations, which relate the geometric properties of spacetime to its matter--energy content through the stress--energy tensor:
\begin{equation}
R_{\mu\nu}-\frac{1}{2}g_{\mu\nu}R = 8\pi T_{\mu\nu},
\label{eq:EFE}
\end{equation}
where $R_{\mu\nu}$ is the Ricci tensor, $g_{\mu\nu}$ is the metric tensor, $T_{\mu\nu}$ denotes the stress--energy tensor,  and $R$ is the Ricci scalar which is formulated in next subsection. 
After some algebra, the Einstein equations are reduced to the following form: Conventionally, the right-hand side of the Einstein equations is attributed to a matter distribution characterized by the energy density \(\rho(r)\), radial pressure \(P_r(r)\), and tangential pressure \(P_t(r)\). The Einstein field equations then read
\begin{equation} \label{g9}
m'(r) = 4\pi r^2 \rho(r)+\frac{Q^2}{2r^2}+\frac{\Lambda r^2}{2}+\frac{\rho_q r^2}{2},
\end{equation}
More specifically, the mass function was reconstructed from the Newtonian relation between the centrifugal and gravitational forces governing circular motion at a radial distance \(r\)
\begin{equation}\label{g10}
\frac{m(r)}{r}=v_{tg}^{2}.
\end{equation}
After reconstructing the mass function \(m(r)\), the metric function \(g(r)\) is obtained directly from Eq.~(\ref{g4}), while the lapse function \(f(r)\) is determined by employing the general relativistic relation for the tangential velocity associated with the same circular orbit
\begin{equation}\label{g11}
v_{tg}^{2}=\frac{rf'(r)}{2 f(r)}.
\end{equation}
Hence,
\begin{equation}\label{g12}
\frac{f'(r)}{f(r)}=\frac{2m(r)}{r^{2}}.
\end{equation}
With the mass function determined from Eq.~(\ref{g10}), the associated density profile follows directly from Eq.~(\ref{g9}).

\subsection{Dehnen-(1,4,5/2) Profile}\label{04}

Another example of the reconstruction approach discussed in Sec.~II is provided in Ref.~\cite{DN1}, where the analysis starts from the Dehnen-((1,4,5/2)) density profile proposed in Ref.~\cite{DN2}. As discussed below, this reconstruction procedure is subject to the same conceptual limitations
\begin{equation}\label{DP1}
\tilde{\rho}(r)=\frac{\rho_s r_s^{4}}{r^{5/2}(r+r_s)^{3/2}} .
\end{equation}
From Eqs. (\ref{g6}), (\ref{g9}) and (\ref{DP1}), one can formulate the metric function as 
\begin{equation}
f(r)=\Psi(r)=1-\frac{2M}{r}-\frac{\Lambda r^{2}}{3}+\frac{Q^{2}}{r^{2}}-\frac{16\pi \rho_s r_s^{3}}{\sqrt{r(r+r_s)}}-{
\frac {C_q}{r^{3\omega_q+1}}}.
\end{equation}
By using this expression $\frac{f'(r)}{f(r)}=\frac{2m(r)}{r^{2}}$, provided as
\begin{small}
\begin{equation}
f(r)=\exp\left[C+\frac{\Lambda r^{2}}{6}-{\frac {c_q}{2 r^{3\omega_q+1}}}+\frac{Q^{2}}{2 r^{2}}-32\pi \rho_s r_s^{2}\sqrt{1+\frac{r_s}{r}}\right].
\end{equation}    
\end{small}
for $C=0$, we have 
\begin{equation}
f(r)=\exp\left[\frac{\Lambda r^{2}}{6}-{\frac {C_q}{r^{3\omega_q+1}}}+\frac{Q^{2}}{2 r^{2}}-32\pi \rho_s r_s^{2}\sqrt{1+\frac{r_s}{r}}\right].
\end{equation}\label{f(r)}

\begin{figure}[h!]
\centering
\includegraphics[width=9cm]{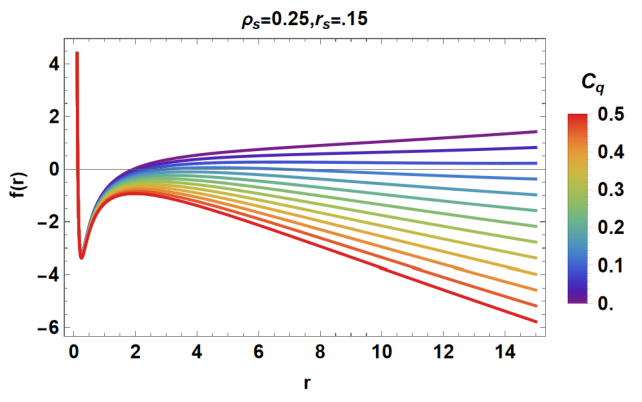}
\caption {\scriptsize Plots of $f(r)$ verses $r$ for Dehnen-(1,4,5/2) profile with different values of quintessence field.} \label{eh1}
\end{figure}
The graphical behavior of the lapse function verses radial coordinate for different choices of quintessence field is presented in Fig. (\ref{eh1}). The position of horizon structure greatly depends on the quintessence field. 

\subsection{NGC 4649 Halo}\label{03}

In Ref.~\cite{NGC1}, two black hole metrics, referred to as the ``first solution'' and the ``second solution'', were proposed for the dark matter halo of NGC~4649 (M60)~\cite{NGC2}, characterized by the density profile
\begin{equation} \label{g13}
\rho(r)=\frac{V_c^{2}}{4\pi}\frac{3a^{2}+r^{2}}{(r^{2}+a^{2})^{2}}.
\end{equation}
where $a$ denotes the characteristic core radius of the halo,
and $V_c$ is the critical tangential velocity associated with
the DM distribution. The spacetime geometry is constructed following the methods outlined in \cite{NGC3, NGC4}, while Observational constraints on the shadow shape are incorporated using results from \cite{NGC5}. In the case of the second solution, the mass function is determined by integrating the following equations: (\ref{g6}), (\ref{g9}) and (\ref{g13}), which yields the metric functions as
\begin{small}
\begin{eqnarray} \nonumber
f(r)=\Psi(r)=1-\frac{2M}{r}-{\frac{2 {V_c}^{2}{r}^{2}}{{a}^{2}+{r}^{2}}}-{\frac{\Lambda{r}^{2}}{3}}+{\frac {{Q}^{2}}{r^2}}-{
\frac {C_q }{r^{3\omega_q+1}}}.
\\
\label{NGCaa}
\end{eqnarray}
\end{small}
\textcolor{black}{
which is the black hole solution of the Einstein equations with the density expression given in Eq.~(\ref{g13}). In this regime, the above solution describes a black hole mass embedded in a dark-matter halo. 
For the limiting case $C_q \to 0$, $\rho_s \to 0$, $Q \to 0$, and $\Lambda \to 0$, the resulting metric reduces to the well-known 
Schwarzschild black hole solution \cite{KS}. For $C_q \to 0$, 
$\rho_s \to 0$, and $\Lambda \to 0$, it recovers the  Reissner--Nordstr\"om (RN) black hole solution \cite{RN1,RN2}.  Furthermore, for $C_q \to 0$ and $\rho_s \to 0$, the metric reduces  to the Reissner--Nordstr\"om--AdS (RN--AdS) black hole solution  \cite{RNADS1,RNADS2}. Finally, in the limit $\Lambda \to 0$ and  $\rho_s \to 0$, the solution reduces to the well-known charged black  hole surrounded by quintessence \cite{Q1,Q2}.  The Ricci scalar \(R\) is a fundamental quantity for characterizing static, spherically symmetric spacetimes. In modified black hole analysis, it provides a convenient diagnostic of the spacetime geometry, allowing us to determine whether the metric is Ricci-flat, identify possible curvature singularities, and verify the consistency of the metric with the corresponding gravitational field equations. So that $R$ specifies the Ricci scalar which is formulated as \cite{CK1,CK2}
\begin{equation}\label{eq:Ricci-general}
R=-f''(r)-\frac{4f'(r)}{r}+\frac{2[1-f(r)]}{r^2}.
\end{equation}
After straightforward algebraic simplification, the Ricci scalar takes the form
\begin{equation}
R=4\Lambda+\frac{4V_c^2\left(6a^4+3a^2r^2+r^4\right)}{\left(a^2+r^2\right)^3}+
\frac{3C_q\omega_q\left(3\omega_q-1\right)}{r^{3\omega_q+3}}.
\label{eq:Ricci-scalar}
\end{equation}
As a consistency check, the contributions associated with the mass parameter $M$ and the electric charge $Q$ vanish identically from the Ricci scalar $R$, as expected for the Schwarzschild and Maxwell sectors. Moreover, in the limiting case $V_c=C_q=0$, one obtains $R=4\Lambda$, which consistently recovers the standard Reissner-Nordstr\"om-(A)dS spacetime.}

\begin{figure}[h!]
\centering
\includegraphics[width=9cm]{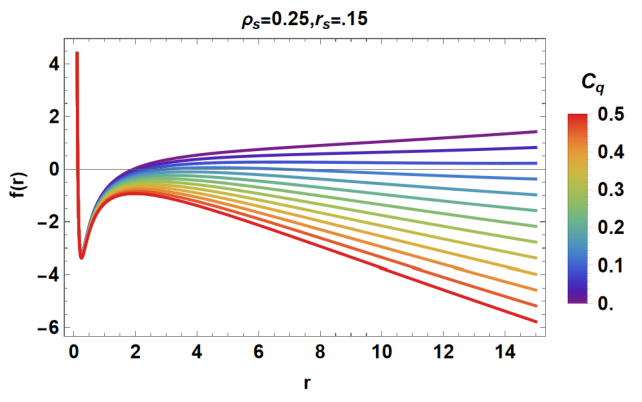}
\caption {\scriptsize Plots of $f(r)$ verses $r$ for NGC 4649 halo with different values of quintessence field.} \label{eh2}
\end{figure}

The plot of the lapse function of NGC 4649 halo verses radial coordinate for different choices of quintessence field is presented in Fig. (\ref{eh2}). We observe the impact of quintessence field on the position of the horizon radius. The Horizon radius moves away from the center of black hole by increasing the quintessence field. 

\section{Smarr relation}\label{05}
In black hole thermodynamics, the thermodynaic valoume $V$ is related to the thermodynamic pressure $P$ and the black hole entropy at the horizon becomes $S= {\pi r_{h}^2}$ as previously defined in Eq. \ref{S} with the event horizon radius $r_h$ is obtained from the condition $f(r_h)=0.$ \textcolor{black}{while the black hole mass $M$  in Eq. (\ref{NGCaa}) in terms of entropy takes the form
\begin{eqnarray} \nonumber \label{M}
M&=&\frac{1}{2}\bigg[\sqrt{\frac{S}{\pi}}+\frac{8PS^{3/2}}{3\sqrt{\pi}}+\frac{Q^{2}\sqrt{\pi}}{\sqrt{S}}-\frac{2V_c^{2}\left(\frac{S}{\pi}\right)^{3/2}}{a^{2}+\frac{S}{\pi}}\\
&-&\frac{C_q}{\left(\frac{S}{\pi}\right)^{\frac{3\omega_q}{2}}}\bigg],
\end{eqnarray}
the halo core radius \(a\) and the quintessence parameter \(C_q\) must be treated consistently. The extended first law of thermodynamics (where halo core radius \(a\) held fixed) is defined as
\begin{equation}
dM=T\,dS+V\,dP+\Phi\,dQ+\mathcal{C}\,dC_q+X\,dV_c,
\label{firstlaw}
\end{equation}
with $V$ be the thermodynamic volume is given by
\begin{equation}
    V=\left(\frac{\partial M}{\partial P}\right)_{S,Q,...}=\frac{4S^{3/2}}{3\sqrt{\pi}}
\end{equation}
while the Hawking temperature $T$ is expressed as
\begin{eqnarray} \nonumber
T=\left(\frac{\partial M}{\partial S}\right)_{Q,P,...}&=&\frac{1}{4\pi}\bigg[\frac{2M}{r_h^{2}}-\frac{2Q^{2}}{r_h^{3}}
-\frac{4V_c^{2}a^{2}r_h}{(a^{2}+r_h^{2})^{2}}\\
&-&\frac{2\Lambda r_h}{3}+\frac{(3\omega_q+1)C_q}{r_h^{3\omega_q+2}}\bigg].
\label{temperature}
\end{eqnarray}
The electric potential is
\begin{equation}
\Phi_e=\left(\frac{\partial M}{\partial Q}\right)_{S,P,...}=\frac{Q}{r_h}=Q\sqrt{\frac{\pi}{S}},
\label{phi}
\end{equation}
so that $Q={\Phi_e}{r_h}$. Treating the quintessence parameter $C_q$ as a thermodynamic variable, its conjugate quantity is as follows
\begin{equation}
\mathcal{C}=\left(\frac{\partial M}{\partial C_q}\right)_{S,P,Q,...}=-\frac{1}{2r_h^{3\omega_q}}=-\frac{1}{2}\left(\frac{\pi}{S}\right)^{\frac{3\omega_q}{2}}.
\label{cq}
\end{equation}
Similarly, if $V_c$ is regarded as a thermodynamic variable, its conjugate quantity for the halo parameter \(V_c\) becomes
\begin{equation}
X=\left(\frac{\partial M}{\partial V_c}\right)_{S,P,Q,...}=-\frac{2V_c r_h^{3}}{a^{2}+r_h^{2}}.
\label{x}
\end{equation}
Applying Euler's scaling theorem leads to the Smarr relation
\begin{equation}
M=2TS-2PV+\Phi_e Q+(1+3\omega_q)\mathcal C C_q+\mathcal A a.
\label{smarr}
\end{equation}
where \(\mathcal A\) be the conjugate to the halo core radius \(a\)
\begin{equation}
\mathcal A=\left(\frac{\partial M}{\partial a}\right) _{S,P,Q,C_q,V_c} =\frac{2aV_c^2r_h^3}{(a^2+r_h^2)^2}
\end{equation}
The first law and Smarr relation provides a consistent framework for studing charged AdS black holes surrounded by the dark matter halo and quintessence, which gives important results about the black hole thermodynamics effected by the environmental matter fields before discussing the geometric modifications. }

\section{Canonical Ensemble}\label{06}
In the following sections, we consider the NGC 4649 halo profile to explore the topological structure. For the quintessence black hole geometry considered in this work, the metric function is expressed in (\ref{NGCaa}), where \(M\) is the black hole mass, \(Q\) denotes the electric charge, and \(\Lambda\) is the cosmological constant. The parameter \(C_q\) characterizes the intensity of the surrounding quintessence field, while \(\omega_q\) represents the quintessence state parameter. The quintessence contribution modifies the asymptotic structure and thermodynamic behavior of the black hole. The mass parameter \(M\) of the black hole derived from Eq. (\ref{M}) is expressed as
\begin{equation}\label{M1}
M=\frac{1}{6} \left[ \frac{3Q^2}{r_h} + 3r_h - 3C_q r_h^{-3\omega_q} + r_h^3 \left( 8\pi P - \frac{6V_c^2}{a^2 + r_h^2} \right) \right].
\end{equation}
The Hawking temperature derived from mass parameter $M$ is written as
\begin{small}
\begin{equation}\label{TH}
T_H = \frac{1}{4 \pi r_h} \left[{1 - \frac{Q^2}{r_h^2} + 2 r_h^2 \left(4 P \pi - X_1\;V_c^2 \right) + 3 C_q r_h^{-1 - 3 \omega_q} \, \omega_q}\right].
\end{equation}
\end{small}
Here, $X_1$ is defined as
\begin{equation}\label{d}
   X_1 =  \frac{\left(3 a^2 + r_h^2\right) }{\left(a^2 + r_h^2\right)^2}.
\end{equation} 
The critical pressure is obtained by imposing the condition that the Hawking temperature attains an extremum with respect to the event horizon radius \(r_h\). In the framework of extended black hole thermodynamics, the occurrence of a second-order phase transition is characterized by the existence of an inflection point in the temperature profile. This criterion is implemented through the conditions $\left({\partial T}/{\partial r_h}\right)_{Q}=0,  \left({\partial^{2}T}/{\partial r_h^{2}}\right)_{Q}=0.$ Solving these equations simultaneously yields the critical horizon radius \(r_c\) together with the corresponding critical pressure \(P_c\). This procedure follows the standard methodology employed in black hole chemistry for identifying second-order phase transitions and is fully consistent with the inflection point analysis discussed in \cite{TC13,TC14,TC15}. Therefore, the critical pressure is not introduced as an arbitrary parameter; rather, it emerges naturally from the thermodynamic requirement that the Hawking temperature exhibits an extremum at the critical point. Consequently, the critical pressure obtained through this analysis can be expressed as
\begin{small}
\begin{equation}\label{P}
P = \frac{1}{8\pi} {\left[- \frac{3Q^2}{r_h^4} + \frac{1}{r_h^2}+{2\; X_2 V_c^2} + 3C_q r_h^{-3(1+\omega_q)} \omega_q (2 + 3\omega_q)\right]}.
\end{equation}
\end{small}
where $X_2$ is defined as
\begin{equation}\label{011}
X_2=\frac{(3a^4-6a^2 r_h^2-r_h^4)}{(a^2+r_h^2)^3}.
\end{equation}
By substituting this result into the temperature relation (\ref{TH}), the corresponding isobaric temperature $T$ becomes
\begin{equation}\label{TP}
T_p = \frac{1}{4 \pi r_h}\left[{2 - \frac{4 Q^2}{r_h^2} + 9 C_q r_h^{-1 - 3 \omega_q} \, \omega_q \left(1 + \omega_q\right)- X_3\; V_c^2 }\right],
\end{equation}
where $X_3$ is defined as
\begin{equation}
X_3 = \frac{4 \left(5 a^2 r_h^4 + r_h^6\right)}{\left(a^2 + r_h^2\right)^3}.
\end{equation} 
Now, we employ the method introduced earlier to analyze the parameter space and determine the zero points of the vector field. These zero points correspond directly to the black hole solutions. The associated topological charge can then be evaluated using $\phi$-mapping and topological current theory. Here, $\phi$ denotes the thermodynamic function and is defined as $\phi = 1/ sin\theta \;T$,
\begin{small}
\begin{equation}
\phi=\frac{1}{4 \pi r_h}{\left[2 - \frac{4 Q^2}{r_h^2} + 9 C_q r_h^{-1 - 3 \omega_q} \, \omega_q \left(1 + \omega_q\right)- X_3 V_c^2 \right]\csc\theta}.
\end{equation}
\end{small}
 The vector field's components  \(\Phi = (\Phi^{r_h}, \Phi^{\theta})\) are 
\textcolor{black}{\begin{eqnarray}\nonumber
    \Phi^{r_h} &=&\frac{1}{4 \pi r_h^2}\bigg[-2  - 9 C_q r_h^{-1 - 3 \omega_q} \, \omega_q \bigg(2 + 5 \omega_q + 3 \omega_q^2\bigg)\\
&-& \frac{12 Q^2}{r_h^2} + X_4\; V_c^2\bigg]\csc{\theta},
\end{eqnarray}}

where $X_4$ is defined as 
\begin{equation}
    X_4= \frac{4 r_h^4 \left(-15 a^4 + 10 a^2 r_h^2 + r_h^4\right) }{\left(a^2 + r_h^2\right)^4}.
\end{equation}
and
\begin{eqnarray} \nonumber
\Phi^\theta &=& -\frac{1}{4 \pi r_h}\bigg[2 - \frac{4 Q^2}{r_h^2} + 9 C_q r_h^{-1 - 3 \omega_q} \, \omega_q \left(1 + \omega_q\right)\\
&-& X_3\; V_c^2 \bigg]\cot{\theta}\,\csc{\theta}.
\end{eqnarray}

The normalized vector field is obtained by dividing each component of the thermodynamic vector field by its magnitude $(\|\Phi\|)$. This procedure preserves the direction of the vector field while fixing its norm to unity, thereby allowing the analysis to focus solely on the geometric and topological structure of the thermodynamic phase space. The normalized components $(\Phi_{r_h}/\|\Phi\|)$ and $(\Phi_{\theta}/\|\Phi\|)$ describe the orientation of the vector field in the $((r_h,\theta))$ plane and are independent of the overall scale of $(\Phi)$. The normalized vector's component expressed as
\begin{small}
\begin{widetext}
\begin{equation}\label{8}
\frac{\Phi^{r_h}}{\left\| \Phi \right\|} =\frac{-2 - \frac{12 Q^2}{r_h^2} - 9 C_q r^{-1 - 3 \omega_q} \, \omega_q \left(2 + 5 \omega_q + 3 \omega_q^2\right) + X_4 V_c^2}{\left[\left(2 + \frac{12 Q^2}{r_h^2} + 9 C_q r_h^{-1 - 3 \omega_q} \omega_q \left(2 + 5 \omega_q + 3 \omega_q^2\right)-X_4 V_c^2\right)^2+ r_h^2 \left(2 - \frac{4 Q^2}{r_h^2} + 9 C_q r_h^{-1 - 3 \omega_q} \omega_q \left(1 + \omega_q\right) - X_3 V_c^2\right)^2 \cot^2{\theta}\right]^\frac{1}{2}},
\end{equation}
\end{widetext}
\end{small}
and
\begin{small}
\begin{widetext}
\begin{equation}\label{9}
\frac{\Phi^\theta}{\left\| \Phi \right\|}=-\frac{r_h\left(2 - \frac{4 Q^2}{r_h^2} + 9 C_q r_h^{-1 - 3 \omega_q} \omega_q \left(1 + \omega_q\right)- X_3 V_c^2\right)\cot{\theta}}{\left[{{\left(2 + \frac{12 Q^2}{r_h^2} + 9 C_q r_h^{-1 - 3 \omega_q} \omega_q \left(2 + 5 \omega_q + 3 \omega_q^2\right)-X_4 V_c^2\right)^2+ r_h^2 \left(2 - \frac{4 Q^2}{r_h^2} + 9 C_q r_h^{-1 - 3 \omega_q}\omega_q \left(1 + \omega_q\right) - X_3 V_c^2\right)^2 \cot^2{\theta}
}}\right]^\frac{1}{2}}.
\end{equation}
\end{widetext}
\end{small}

\begin{figure}[h!]
\centering
\includegraphics[width=8cm]{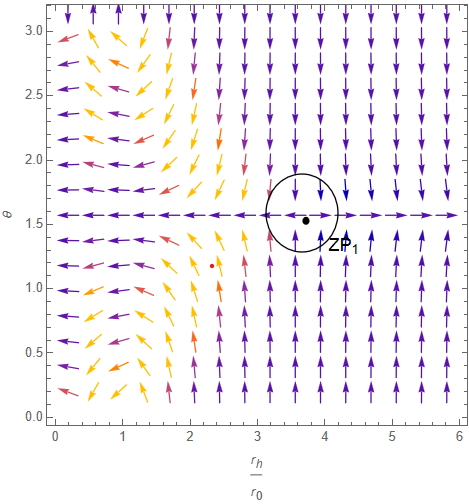}
\caption {\scriptsize Normalized vector field $n$ plotted in the $(r_h/r_0,\theta)$ plane for the black hole system. The dot indicates the critical point.} \label{fig1}
\end{figure}

The normalized vector is defined as $\vec{n} = \left(\frac{\Phi_r}{\| \Phi \|}, \frac{\Phi_\theta}{\| \Phi \|} \right)$ and its behavior in the \( r_h \) vs \( \theta \) plane for the black hole is illustrated in Fig. \ref{fig1}. The plot is generated using the parameter values  \( \omega_q=-0.6\), \(a=10\), \(V_c=0.005\), \(Q=0.1\) and \(C_q=0.90\). The black dot in the figure marks the critical point, denoted by  $ZP_1$. To determine this critical point, one sets \( \theta = \tfrac{\pi}{2} \) in Eq. (\ref{8}) for zero point $ZP_1$ and solves the condition under which the expression vanishes. The behavior of the surrounding vectors shows a structured flow pattern around these points, suggesting that nearby trajectories are significantly influenced by their presence.

\begin{figure}[h!]
\centering
\includegraphics[width=0.5\textwidth]{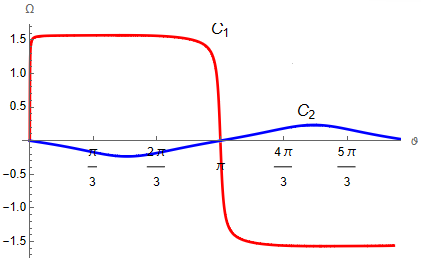}
\caption {\scriptsize Variation of $\Omega$ with respect to $\vartheta$ along the contours $C_1$ and $C_2$, illustrating the angular dependence of $\Omega$ across different
trajectories.}\label{O1}
\end{figure}

Fig. \ref{O1} illustrates the angular dependence of $\Omega(\vartheta)$ over the domain $\vartheta \in [0, 2\pi]$ for two distinct contours with \((a, b, r_{0}) =
(0.993, 1.495, 3.97)\) for \(C_{1}\) and \((a, b, r_{0}) = (0.05, 0.998, 0.5)\) for \(C_{2}\). The red curve, corresponding to $C_1$, exhibits a nearly monotonic profile, starting from a positive value at $\vartheta = 0$, then undergoing a sharp transition around $\vartheta = \pi$, where it rapidly crosses zero and approaches a negative constant as $\vartheta \to 2\pi$. This abrupt change signifies the presence of a critical angular point and indicates a nontrivial topological transition. The overall variation of $\Omega(\vartheta)$ along $C_1$ results in a net phase change, yielding a winding number of $-1$. In contrast, the blue curve, representing $C_2$, shows a smooth and oscillatory behavior, remaining finite and continuous throughout the interval and returning to its initial value at $\vartheta = 2\pi$. Since this oscillatory mode completes a full cycle without any net phase accumulation, it does not contribute to the topological charge. Consequently, the topological charge associated with $C_1$ is $W_{C_1} = -1$, while for $C_2$ it is $W_{C_2} = 0$, leading to a total topological charge $W = -1$.

\section{Canonical Ensemble as black hole Solutions}\label{07}
Substituting the black hole mass and entropy into the general expression for the generalized free energy renders this quantity especially useful for exploring thermodynamic properties, such as phase transitions and critical behavior. Using the mass from Eq. (\ref{M}) together with the entropy of the black hole from Eq. (\ref{S}), the generalized free energy (\ref{F0}) is obtained as:
\begin{small}
\begin{equation}
F = \frac{1}{6} \left[\frac{3 Q^2}{r_h} + 3 r_h + r_h^3 \left(8 \pi P-\frac{6 V_c^2}{a^2 +r_h^2}\right)\right] - 3 C_q r_h^{-3 \omega_q} -\frac{\pi r_h^2}{\tau}.
\end{equation}
\end{small}
The components of vector fields defined in Eq. (\ref{5}) are calculated as
\begin{small}
\begin{equation}
\Phi^{r_h}=\frac{1}{2} \bigg[1 - \frac{Q^2}{r_h^2}- \frac{4 \pi r_h}{\tau}+ 3 C_q r_h^{-1-3 \omega_q}\omega_q
+r_h^2 \left(8 \pi P-2 X_1 V_c^2 \right)\bigg],    
\end{equation}
\end{small}
here $X_1$ is defined in Eq. (\ref{d}) and $\Phi^\theta$ gives in Eq. (\ref{5}).
\begin{eqnarray} \nonumber 
\| \Phi \|&=&\bigg[\frac{1}{4} \bigg(1 - \frac{Q^2}{r_h^2}- \frac{4 \pi r_h}{\tau}+3 C_q r_h^{-1 - 3 \omega_q}  \omega_q\\
&+& r_h^2 (8 \pi P -2 X_1\;V_c^2 )\bigg)^2+\cot^2{\theta}\csc^2{\theta} \bigg]^\frac{1}{2}.   
\end{eqnarray}
The normalized vectors ensuring that the resulting vector field has unit norm. This normalization removes the dependence on the vector magnitude and retains only its directional information. In thermodynamic topology, the normalized vectors are used to investigate the winding behavior of the vector field around its zero points, which correspond to thermodynamic critical states. The orientation of the normalized vector field determines the topological charge and provides a powerful tool for characterizing phase transitions and the global thermodynamic structure of the black hole.The normalized vectors are expressed as

\begin{widetext}
\begin{equation}
\frac{\Phi^{r_h}}{\left\| \Phi \right\|}=\frac{\frac{1}{2} \left(1 - \frac{Q^2}{r_h^2}+  r_h^2 \left(8 \pi P - 2 X_1\;V_c^2 \right)- \frac{4 \pi r_h}{\tau}+ 3 C_q r_h^{-1 - 3 \omega_q} \, \omega_q\right)}{\left[{\frac{1}{4} \left(1 - \frac{Q^2}{r_h^2}+  r_h^2 \left(8 \pi P -2 X_1V_c^2 \right)- \frac{4 \pi r_h}{\tau}+ 3 C_q r_h^{-1-3\omega_q} \omega_q\right)^2+\cot^2{\theta}\csc^2{\theta}}\right]^\frac{1}{2}},
\end{equation}
\end{widetext}
and
\begin{widetext}
\begin{equation}
\frac{\Phi^{\theta}}{\left\| \Phi \right\|}=-\frac{\cot{\theta}\,\csc{\theta}}{\left[{\frac{1}{4} \left(1 - \frac{Q^2}{r_h^2}+ r_h^2 \left(8 \pi P -2 X_1\;V_c^2 \right)- \frac{4 \pi r_h}{\tau}+ 3 C_q r_h^{-1 - 3\omega_q}\,\omega_q\right)^2+\cot^2{\theta}\,\csc^2{\theta}}\right]^\frac{1}{2}}.
\end{equation}
\end{widetext}

\begin{figure}[h!]
\centering
\includegraphics[width=8cm]{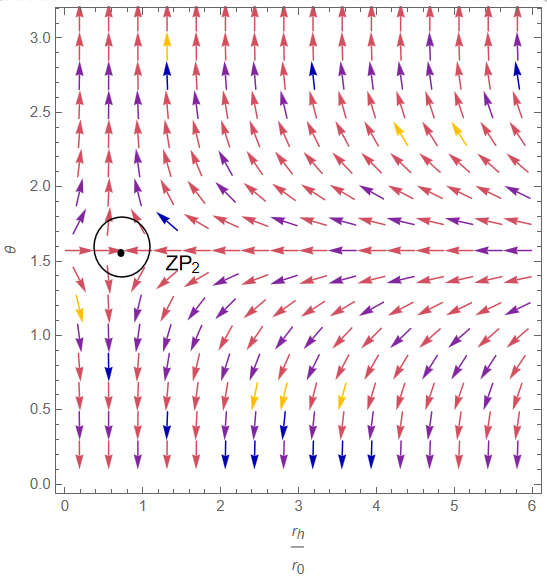}
\caption {\scriptsize Normalized vector field $n$ plotted in the $(r_h/r_0,\theta)$ plane for the black hole system. The dot indicates the critical point.} \label{fig2}
\end{figure}
Fig. \ref{fig2} represents the phase space structure of the system, where the vector field describes the dynamical flow. The black dot in the figure marks the zero point, labeled as $ZP_2$, which corresponds to a critical point where both components of the vector field vanish. This point is obtained by solving the condition under which the field components become zero simultaneously. In this way, the location of the critical point is given by $(r_h, \theta) \approx (0.40, \pi/2)$ for the parameter values \( \omega_q=-0.6\), \(a=10\), \(V_c=0.005\), \(Q=0.1\), \(P=0.005\), \(C_q=0.94\) and \(\tau=10\), respectively.

\begin{figure}[b!]
\centering
\includegraphics[width=8cm]{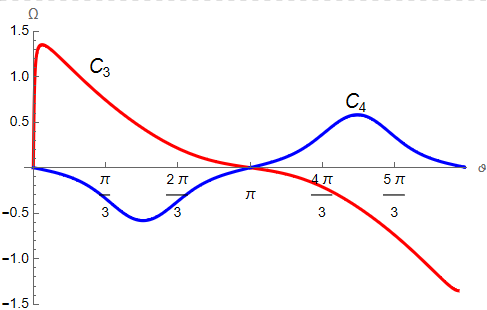}
\caption {\scriptsize Variation of $\Omega$ with respect to $\vartheta$ along the contours $C_3$ and $C_4$, illustrating the angular dependence of $\Omega$ across different
trajectories.}\label{O2}
\end{figure}
Fig. \ref{O2} illustrates the angular dependence of $\Omega(\vartheta)$ over the interval $\vartheta \in [0,2\pi]$ for $C_3$ (red) with \((a, b, r_{0}) = (0.10,0.25, 0.176)\) and $C_4$ (blue) with \((a, b, r_{0}) = (0.14,1.10, 1.68)\). The red curve ($C_3$) starts from a positive value, reaches a local maximum (critical point), and then decreases, crossing zero and becoming negative, indicating a net phase change and a nontrivial topological structure with winding number $W_{C_3} = -1$. In contrast, the blue curve ($C_4$) begins with negative values, attains a local minimum (critical point), then increases to cross zero and form a positive peak before returning toward its initial state, completing a full cycle without net phase accumulation, thus giving $W_{C_4} = 0$. The zero crossings represent points where $\Omega=0$, while the extrema correspond to critical points together determine the qualitative behavior and topological nature of the system. Consequently, the topological charge associated with $C_3$ is $W_{C_3} = -1$, while for $C_4$, it is $W_{C_4} = 0$, leading to a total topological charge $W = -1$.

While the Jacobian method provides an algebraic determination of the winding number, the contour mapping offers a more intuitive geometric interpretation of the thermodynamic topology \cite{CC}. Closed contours enclosing the zero point in the $(r_h,\theta)$ plane are mapped into the $(\phi^{r_h},\phi^\theta)$ plane, where they form image curves $\Phi_i$. The winding of each image curve around the origin determines the corresponding topological charge, with counterclockwise (clockwise) rotation giving $w=+1$ ($w=-1$). Since the winding number is a topological invariant, it remains unchanged under continuous deformation of the contour, provided no additional zero points are enclosed. The winding numbers are first determined analytically from the sign of the Jacobian determinant of the thermodynamic vector field at each zero point. To provide a geometric interpretation and verify the analytical results, closed contours enclosing the zero points are mapped into the $(\phi^{r_h},\phi^\theta)$ plane, where the number and orientation of the rotations around the origin reproduce the same topological charges.

\begin{figure}[hb]
\centering
\includegraphics[width=8cm]{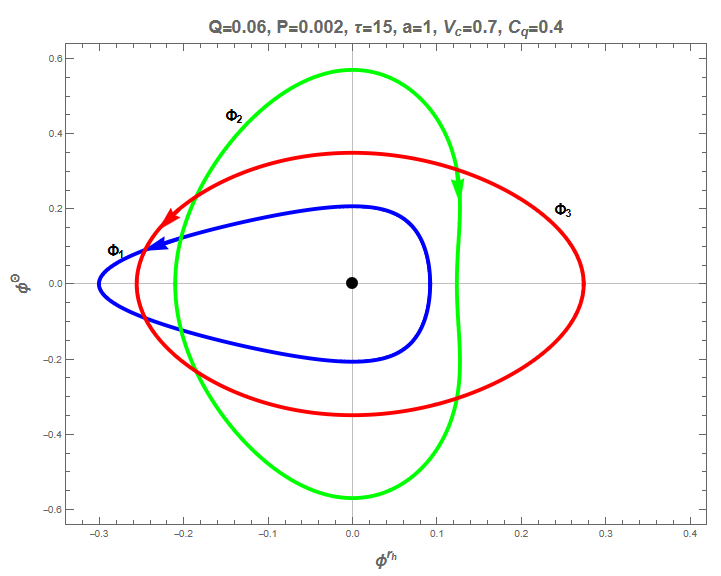}
\caption {\scriptsize Closed curves $\Phi_i$ are obtained as the images of contours $C_i$ enclosing zero points in the $(r_h,\theta)$-plane. The direction of winding of $\Phi_i$ about the origin encodes the topological charge, where counterclockwise winding yields $W=+1$ and signifies thermodynamic stability, while clockwise winding gives $W=-1$, indicating an unstable black hole phase.}\label{cc1}
\end{figure}
In Fig. \ref{cc1}, the topological phase portrait in the $(\Phi^{r_h},\Phi^{\theta})$ plane is shown for the fixed parameter values $Q=0.06$, $P=0.002$, $\tau=15$, $a=1$, $V_c=0.7$, and $C_q=0.4$. The black dot located at the origin represents the zero (critical) point of the thermodynamic vector field, around which three closed contours, $\Phi_1$, $\Phi_2$, and $\Phi_3$, are constructed to evaluate the topological charge. Each contour encloses the same critical point while following a distinct path in the vector field space, as indicated by the arrows. According to the Duan $\phi$-mapping topological current theory, the winding of these contours around the origin determines the corresponding topological charge. The orientation of each contour determines its winding number: a counterclockwise (clockwise) encirclement corresponds to a positive (negative) topological charge, whereas a contour that does not complete a net winding around the origin possesses a vanishing topological charge. Consequently, the winding number serves as a topological invariant that distinguishes stable and unstable black hole phases independently of the contour's shape or size, highlighting the topological robustness of the thermodynamic classification.
This geometric construction offers a clear and intuitive visual verification of the topological charges associated with each black hole solution. 

\section{Mixed Ensembles}\label{08}
For the mixed ensemble of black hole thermodynamics, the electric potential \( \phi_e \) persist constant which is expressed as $\phi_e = {Q}/{r_h}$. The modified mass according to the above assumption is
\begin{small}
\begin{equation}
M=\frac{1}{6} \left[-3 C r_h^{-3 \omega_q}+ r_h^3 \left(8 \pi P - \frac{6 V_c^2}{a^2 + r_h^2}\right)+ 3 r_h \left(1 + \phi_e^2\right)\right],  
\end{equation}    
\end{small}
and the modified temperature is
\begin{small}
\begin{equation}
    T=\frac{1}{4 \pi r_h}\left[{1 + r_h^2 \left(8 \pi P - 2 X_1\;V_c^2 \right)+ \phi_ e^2+ 3 C_q r_h^{-1 - 3 \omega_q} \, \omega_q}\right].
\end{equation}    
\end{small}
where $X_1$ is defined in Eq. (\ref{d}). Thermodynamics function is $\phi = {1}/{\sin\theta}\;T$, and the vector field \( \boldsymbol{\Phi} = (\Phi^{r_h}, \Phi^{\theta}) \) has the following vector components which are written as
\begin{eqnarray} \nonumber
    \Phi^{r_h}&=&\frac{1}{4 \pi}\bigg[8 P \pi- 3 C r_h^{-3 (1 + \omega_q)} \, \omega_q \left(2 + 3 \omega_q\right) + 2\; X_2 V_c^2 \\
    &+&\frac{-1 - \phi e^2}{r_h^2}\bigg]\csc{\theta},
\end{eqnarray}
where $X_2$ is defined in Eq. (\ref{011}) and

\begin{eqnarray} \nonumber
\Phi^{\theta} &=&- \frac{1}{4 \pi r_h}\bigg[ r_h^2 \left(8 \pi P - 2 X_1 V_c^2 \right) +  3 C_q r_h^{-1 - 3 \omega_q} \, \omega_q\\
&+&1 + \phi_e^2\bigg]\cot{\theta}\,\csc{\theta}.
\end{eqnarray}
These vectors characterize the local orientation of the thermodynamic flow in the $((r_h,\theta))$ space and are essential for analyzing the behavior of the vector field near its zero points. In the thermodynamic topology framework, they are used to determine winding numbers and topological charges associated with black hole phase transitions.The normalized vector components are
\begin{small}
\begin{widetext}
\begin{equation}
\frac{\Phi^{r_h}}{\left\| \Phi \right\|}=\frac{r_h\left(-8 \pi P - 2 X_2 V_c^2 + \frac{-1 + \phi e^2}{r_h^2}- 3 C_q r_h^{-3(1 + \omega_q)} \omega_q \left(2 + 3 \omega_q\right)\right)}{\left[{r_h^2\left(8 \pi P + 2 X_2V_c^2 + \frac{1 - \phi e^2}{r_h^2}+3C_q r_h^{-3(1 + \omega_q)}\omega_q \left(2 + 3 \omega_q\right)\right)^2+\left(1 + r_h^2 \left(-8 \pi P + 2 X_1 V_c^2 \right)- \phi e^2+ 3 C_q r_h^{-1 - 3 \omega_q} \omega_q\right)^2 \cot^2{\theta}}\right]^\frac{1}{2}},
\end{equation}
\end{widetext}
\end{small}
and
\begin{small}
\begin{widetext}
\begin{equation}
\frac{\Phi^{\theta}}{\left\| \Phi \right\|}= \frac{-\left(1 + r_h^2 \left(-8 \pi P + 2 X_1 V_c^2 \right)- \phi e^2+ 3 C_q r_h^{-1 - 3 \omega_q} \, \omega_q\right)\cot{\theta}}{\left[{r_h^2\left(8 \pi P + 2 X_2 V_c^2 + \frac{1 - \phi e^2}{r_h^2}+ 3 C_q r_h^{-3(1 + \omega_q)} \, \omega_q \left(2 + 3 \omega_q\right)\right)^2+ \left(1 + r_h^2 \left(-8 \pi P + 2 X_1 V_c^2 \right)- \phi e^2+ 3 C_q r_h^{-1 - 3 \omega_q} \, \omega_q\right)^2 \cot^2{\theta}}\right]^\frac{1}{2}}.
\end{equation}
\end{widetext}
\end{small}

\begin{figure}[b!]
\centering
\includegraphics[width=8cm]{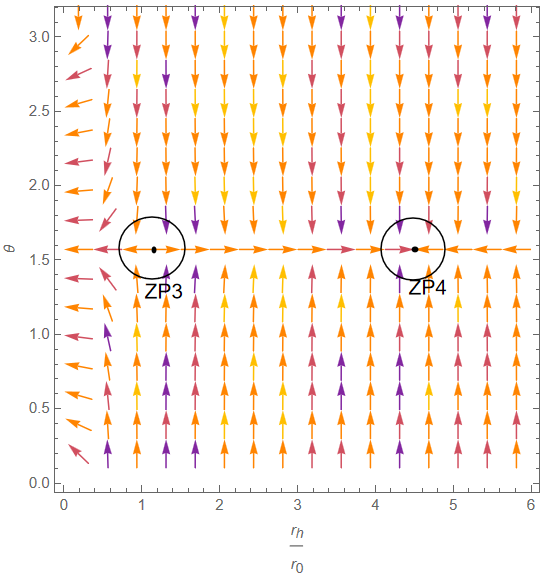}
\caption {\scriptsize Normalized vector field $n$ plotted in the $(r_h/r_0,\theta)$ plane for the black hole system. The dot indicates the critical point.} \label{fig3}
\end{figure}

\begin{figure}[h!]
\centering
\includegraphics[width=8cm]{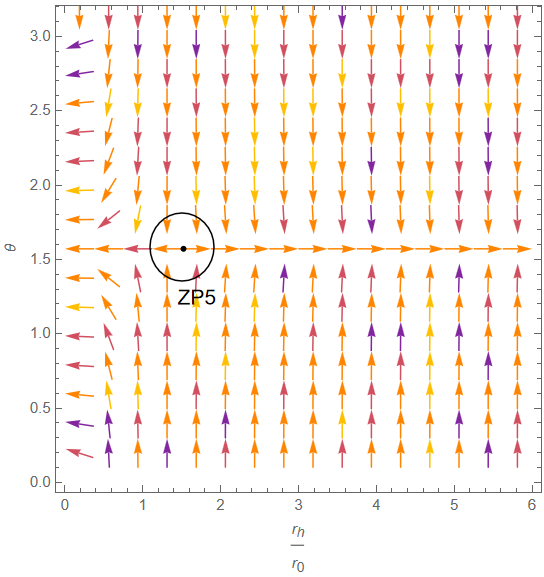}
\caption {\scriptsize Normalized vector field $n$ plotted in the $(r_h/r_0,\theta)$ plane for the black hole system. The dot indicates the critical point.} \label{fig4}
\end{figure}

Fig. \ref{fig3} and \ref{fig4} represents the phase space in the $(r_h/r_0,\theta)$ plane. Fig. \ref{fig3} exhibits two zero points, labeled as $ZP_3$ and $ZP_4$, where the vector field vanishes, indicating that both components of the flow become zero simultaneously. For the parameter values \(C_q=0.90\) and \(\Phi_e=0.10\), these points correspond to critical (equilibrium) configurations of the system and are obtained by solving the conditions under which the vector components vanish. The locations of the critical points are approximately given by $(r_h, \theta) \approx (1.2, \pi/2)$ for $ZP_3$ and $(r_h, \theta) \approx (4.5, \pi/2)$ for $ZP_4$, demonstrating the existence of multiple equilibrium states in the system. Fig. \ref{fig4} exhibit a zero point, labeled as $ZP_5$, which corresponds to a critical point where both components of the vector field vanish. For the parameter values \(C_q=0.97\) and \(\Phi_e=0.10\), these points correspond to critical (equilibrium) configurations of the system and are obtained by solving the conditions under which the vector components vanish. The location of the critical point is approximately given by $(r_h, \theta) \approx (0.52, \pi/2)$ for $ZP_5$, demonstrating the existence of equilibrium state in the system.

\begin{figure}[hb]
\centering
\includegraphics[width=8cm]{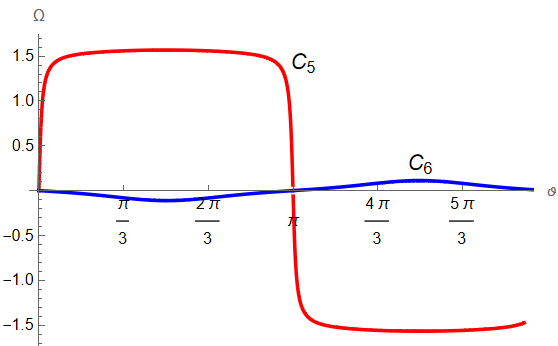}
\caption {\scriptsize Variation of $\Omega$ with respect to $\vartheta$ along the contours $C_5$ and $C_6$, illustrating the angular dependence of $\Omega$ across different
trajectories.}\label{O3}
\end{figure}

Fig.\ref{O3} illustrates the angular dependence of $\Omega(\vartheta)$ over the interval $\vartheta \in [0,2\pi]$ for the curves $C_{5}$ (red) with \((a, b, r_{0}) = (0.895,
1.5, 8.9)\) and $C_{6}$ (blue) with \((a, b, r_{0}) = (0.1,
0.9, 0.19)\). The red curve $C_{5}$ indicates a nontrivial phase transition and reflects a topologically nontrivial structure with winding number $W_{C_{5}}=-1$. In contrast, the blue curve $C_{6}$ exhibits only small smooth oscillations around $\Omega =0$, which represents a topologically trivial configuration with winding number $W_{C_{6}}=0$. The local extrema correspond to critical points that characterize the stable behavior of the system, while the gentle zero crossings indicate weak rotational variation without topological transition. Consequently, the topological charge associated with $C_{5}$ is $W_{C_{5}}=-1$, whereas for $C_{6}$, it is $W_{C_{6}}=0$, giving the total topological charge of the system as $W=-1$.

\section{Flaws in Mixed Ensemble}\label{09}
We next analyze the the thermodynamic topology in the mixed ensemble where the electric potential held fixed. Accordingly, we use the appropriate thermodynamic potential obtained through a Legendre transformation of the fixed-charge description as
\begin{equation} \label{F}
F = M - \phi_e Q - \frac{S}{\tau},
\end{equation}
Substituting values of M from Eq. (\ref{M}) and entropy from Eq. (\ref{S}) in Eq. (\ref{F}).

\begin{eqnarray}\nonumber
F &=&\frac{1}{6} \bigg[- 3 C_q r_h^{-3 \omega_q} + {r_h^3 \left( 8 P \pi - \frac{6 V_c^2}{a^2 + r_h^2} \right)}-\frac{6 \pi r_h^2}{\tau}\\
&+& 3 r_h(1+\phi_e^2) - 6 Q \phi_e \bigg].
\end{eqnarray}
this can be written in the form of potentials as
\begin{eqnarray} \nonumber
\Phi^{r_h}&=&\frac{1}{2} \bigg[1 + 2 r_h^2 \left( 4 P \pi - X_1 V_c^2 \right)+ \phi_e^2 - \frac{4 \pi r_h}{\tau}\\
&+& 3 C_q r_h^{-1 - 3 \omega_q} \, \omega_q\bigg],
\end{eqnarray}
here, $X_1$ is defined in Eq. (\ref{d}) and $\Phi^\theta$ given in Eq. (\ref{5}). These vectors provide a geometric description of the behavior of the system in the thermodynamic parameter space and facilitate the identification of critical points where the vector field vanishes. Within the framework of thermodynamic topology, the normalized vector field is employed to evaluate winding numbers and topological charges, which characterize the stability and phase structure of the black hole and reveal the occurrence of thermodynamic phase transitions. The corresponding normalized vectors are derived as
\begin{widetext}
\begin{equation}
\frac{\Phi^{r_h}}{\left\| \Phi \right\|}=\frac {1 + \phi_e^2 + 2 r_h^2 \left( 4 P \pi - X_1 V_c^2 \right)- \frac{4 \pi r_h}{\tau}+ 3 C_q r_h^{-1 - 3 \omega_q} \, \omega_q}{2\left[{\frac{1}{4} \left(1 + \phi_e^2 + 2 r_h^2 \left( 4 P \pi -X_1 V_c^2 \right)- \frac{4 \pi r_h}{\tau}+ 3 C_q r_h^{-1 - 3 \omega_q} \, \omega_q\right)^2+ \cot^2\theta \, \csc^2\theta}\right]^\frac{1}{2}},
\end{equation}
\end{widetext}
and
\begin{widetext}
\begin{equation}
\frac{\Phi^{\theta}}{\left\| \Phi \right\|}=- \frac{\cot\theta \, \csc\theta}{\left[{\frac{1}{4} \left(1 +\phi_e^2+ 2 r_h^2 \left( 4 P \pi - X_1 V_c^2 \right)- \frac{4 \pi r_h}{\tau}+ 3 C_q r_h^{-1 - 3 \omega_q} \, \omega_q\right)^2+ \cot^2\theta \, \csc^2\theta}\right]^\frac{1}{2}}.
\end{equation}
\end{widetext}

\begin{figure}[h!]
\centering
\includegraphics[width=8cm]{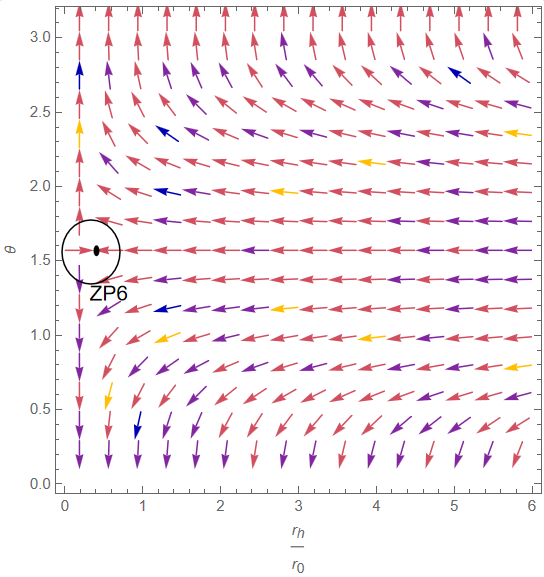}
\caption {\scriptsize Normalized vector field $n$ plotted in the $(r_h/r_0,\theta)$ plane for the black hole system. The dot indicates the critical point.} \label{fig5}
\end{figure}
The zero point is determined by setting the angle \( \theta =
\frac{\pi}{2} \) in \( \frac{\Phi_r}{\left\| \Phi \right\|} \) and
imposing the condition that the expression equals zero. As an
illustration, for chosen parameter values: \(V_c = 0.005 \), \(a = 10 \), \(C_q = 0.2 \) , we obtain a zero point ${ZP_6}$, as shown in Fig. \ref{fig5} at the coordinates $\left( {r_h}/{r_0}, \Theta \right) = (0.40,\frac{\pi}{2}).$
\begin{figure}[h!]
\centering
\includegraphics[width=8cm]{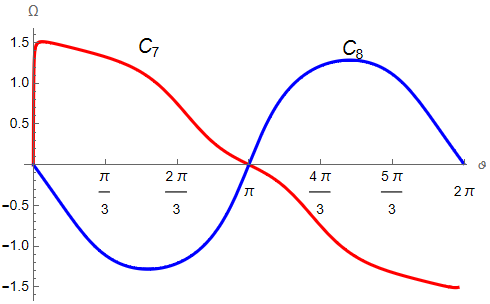}
\caption {\scriptsize Variation of $\Omega$ with respect to $\vartheta$ along the contours $C_7$ and $C_8$, illustrating the angular dependence of $\Omega$ across different
trajectories.}\label{O4}
\end{figure}

Fig. \ref{O4} illustrates the angular dependence of $\Omega(\vartheta)$ for the topological curves $C_{7}$ (red) with \((a, b, r_{0}) = (0.1, 0.49, 0.176)\) and $C_{8}$ (blue) with \((a, b, r_{0}) = (0.1,0.9, 0.6)\) over the interval $\vartheta \in [0,2\pi]$. The red curve $C_{7}$ starts from a positive value and continuously decreases, crossing zero near $\vartheta=\pi$ before reaching a negative value at $2\pi$. This overall phase change indicates a negative circulation around the critical point, corresponding to a winding number $W_{C_{7}}=-1$. In contrast, the blue curve $C_{8}$ initially decreases to a negative minimum, then rises and crosses zero near $\vartheta=\pi$, attaining a positive maximum before returning toward zero at $2\pi$. This behavior reflects a positive phase circulation and corresponds to a winding number $W_{C_{8}}=0$.  Consequently, the topological charge associated with $C_{7}$ is $W_{C_{7}}=-1$, whereas for $C_{8}$, it is $W_{C_{8}}=0$, giving the total topological charge of the system as $W=-1$.

\begin{figure}[h!]
\centering
\includegraphics[width=8cm]{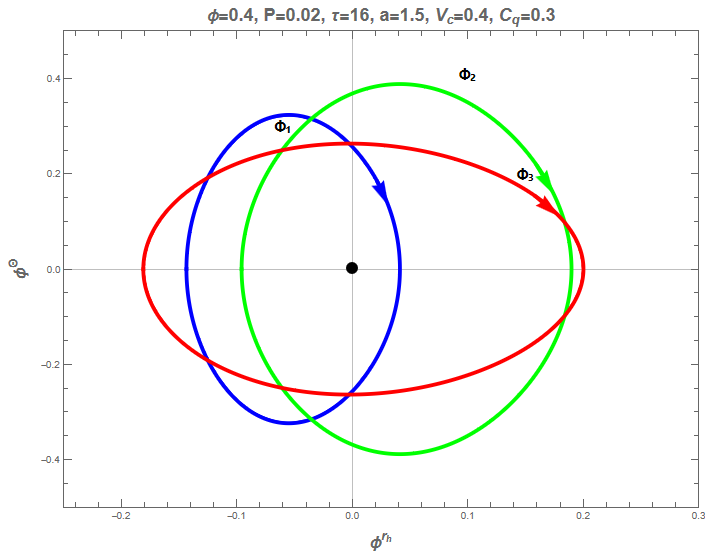}
\caption {\scriptsize Topological phase portrait of the thermodynamic vector field in the $(\phi^{r_h},\phi^\theta)$ plane for $Q=0.5$, $P=0.02$, $\tau=8$, $a=3$, $V_c=0.98$, and $C_q=0.8$. Three closed contours, $\Phi_1$, $\Phi_2$, and $\Phi_3$, with different shapes and sizes enclose the same critical point.}\label{cc2}
\end{figure}

In Fig. \ref{cc2}, the black dot denotes the zero (critical) point of the vector field. The arrows indicate the direction of contour traversal used in the winding number calculation. The figure illustrates that the contours differ considerably in their geometry and enclosed area, they all surround the same critical point. This illustrates one of the fundamental properties of thermodynamic topology: the topological charge depends only on the winding of the vector field around the zero point and not on the particular contour chosen. Therefore, the figure demonstrates the topological invariance of the thermodynamic defect. Consequently, the stability classification of the black hole is determined by the winding number rather than by the contour's size, eccentricity, or orientation in the phase plane. This geometric construction provides clear vision of the topological charges assigned to each black hole solution, thereby reinforcing the consistency of the topological classification. For the Fig. \ref{cc2} phase portrait, all three contours wind clockwise around the origin, yielding  $w(\Phi_1)=w(\Phi_2)=w(\Phi_3)=-1$ and hence $W=-1$, the negative topological charge corresponds to clockwise winding and is associated with a topologically unstable black hole configuration.

\section{Summary and conclusions}\label{010}

In this work, we investigated the topological analysis of charged black holes surrounded by dark matter halos in the presence of a quintessence field. We first reviewed the fundamental properties of quintessence field for dark energy and analyzed its influence on the black hole spacetime geometry. By incorporating both dark matter halo distributions and the quintessence field into the gravitational framework, we derived the corresponding field equations and examined two physically motivated dark matter halo models, namely the NGC 4649 halo in Eq. (\ref{NGCaa}) and the Dehnen-(1,4,5/2) profile in (\ref{f(r)}). Particular attention was given to understanding how the quintessence parameter and dark matter halo characteristics modify the thermodynamic behavior and phase structure of the black hole system. Moreover, the thermodynamic properties of the resulting black hole solutions were explored through the derivation of the Smarr relation in Eq. (\ref{smarr}) and the associated quantities in Eq.s (\ref{temperature}), (\ref{firstlaw}). Incorporating dark matter halos and the quintessence field makes the model more physically consistent by accounting for environmental effects expected around supermassive black holes. The presence of the quintessence field was found to significantly affect the horizon structure and thermodynamic observables, thereby enriching the physical characteristics of the spacetime.

 Furthermore, we examined thermodynamic topology formalism to investigate the topological nature of black hole phase transitions. Within the canonical ensemble, the topological charge and defect structure were analyzed through the behavior of the thermodynamic vector field. The results revealed the existence of topologically stable black hole configurations characterized by the winding numbers (see Fig.s \ref{O1}, \ref{O2}, \ref{O3}, \ref{O4}). We then extended the analysis to mixed canonical ensembles, where additional thermodynamic degrees of freedom produce a richer topological structure and a more intricate critical behavior. The closed contour graphs provide a clear geometric illustration of the thermodynamic topology of the black hole system. In both cases, different closed contours enclosing the same zero point exhibit identical topological behavior, confirming that the winding number and the corresponding topological charge are independent of the contour geometry (see Figs. \ref{cc1}, \ref{cc2}).  The associated winding number and topological charge that classify black hole phases and distinguish stable from unstable thermodynamic configurations. In addition, we compared these results with recently published article in Refs. \cite{TC13,TC14,TC15,TC16,TC17,CC}.

Our results demonstrate that the combined effects of dark matter halos and the quintessence field substantially influence the thermodynamic topology of charged black holes. The topological approach provides a powerful and complementary framework for identifying stable phases and characterizing critical phenomena. These results contribute to a deeper understanding of the interplay between dark matter, dark energy, and black hole thermodynamics, and may offer further insights into the microscopic origin of black hole phase transitions and the role of topological invariants in gravitational systems. Future work may extend this framework to rotating and higher-dimensional black holes, modified gravity theories, and generalized entropy models while exploring their thermodynamic topology. Furthermore, establishing connections between topological invariants and observable signatures, such as black hole shadows, quasinormal modes, gravitational lensing, and gravitational waves, may provide important avenues for testing these predictions through future astrophysical observations.

\section{Numerical Determination of the Critical Points}

In this section, we present the numerical procedure used to determine the thermodynamic critical points of the black hole system in the presence of the quintessence field. The analysis is performed by varying the quintessence state parameter $\omega$ while keeping the remaining black hole parameters fixed. This allows us to investigate how the quintessence field influences the critical radius, critical temperature, and critical pressure.

\subsection{Determination of the AdS Radius}

In the extended phase-space formalism, the cosmological constant is interpreted as the thermodynamic pressure through Eq. (\ref{3}), where $l$ denotes the AdS radius. For a prescribed value of the thermodynamic pressure, the corresponding AdS radius is obtained from $l=\sqrt{{3}/{8\pi P}}.$ Throughout the numerical analysis, the AdS radius is determined from the chosen pressure parameter and is kept fixed while the quintessence parameter $\omega_q$ is varied. Consequently, all variations in the critical quantities originate solely from the influence of the quintessence field rather than changes in the background AdS geometry. For the numerical calculations presented in this work, we choose $P=0.005$, which corresponds to an AdS radius
$l\approx 4.886.$

\subsection{Critical points for different values of $\omega_q$}
The critical quantities $(r_c, T_c, P_c)$ are obtained by solving the standard criticality conditions
\begin{equation}
\left(\frac{\partial P}{\partial r_h}\right)_{T=T_c,\;r_h=r_c}=0,
\qquad
\left(\frac{\partial^2 P}{\partial r_h^2}\right)_{T=T_c,\;r_h=r_c}=0.
\label{eq:critical_conditions}
\end{equation}
The resulting nonlinear equations are solved numerically for fixed values of the remaining black hole parameters while varying the quintessence state parameter $\omega_q$. The corresponding critical radius, critical temperature, and critical pressure are summarized in Table \ref{tab:critical_values}. The numerical results demonstrate that the presence of the quintessence field significantly modifies the thermodynamic critical behavior, leading to systematic variations in the critical quantities as $\omega_q$ changes.\\
\textcolor{black}{\begin{table}[ht]
\centering
\caption{Critical points $(r_c,T_c,P_c)$ for fixed \(Q=0.5\), \(a=10\), \(V_c=0.005\), and \(C_q=0.2\), and varing \(\omega_q\) from \(-0.4\) to \(-0.9\).}
\label{tab:critical_values}
\begin{tabular}{cccc}
\hline\hline
$\omega_q$ & $r_c$ & $T_c$ & $P_c$ \\
\hline
$-0.4$  & 1.355  & 0.058  & 0.0083  \\
$-0.5$ & 1.312 & 0.055 & 0.0090 \\
$-0.6$ & 1.258 & 0.054 & 0.0110 \\
$-0.7$ & 1.211 & 0.056 & 0.0146\\
$-0.8$ & 1.182 & 0.062 & 0.0201\\
$-0.9$ & 1.181 & 0.072 & 0.0275 \\
\hline
\end{tabular}
\end{table}
In Table \ref{tab:critical_values}, as the  $\omega_q$ decreases, the horizon radius also decreases, and the critical pressure increases throughout the range. This represents a substantial increase in the pressure associated with the critical point. The temperature firstly decrease then increase, which show the monotonic behavior indicates that the dependence of the critical temperature on the quintessence parameter is not necessarily monotonous. The results shows that the quintessence parameter has significant effects on the critical behavior of the black hole.}

\subsection{Fractional shifts relative to the reference value of $\omega_q$}

To quantify the influence of the quintessence parameter, the absolute shift of a thermodynamic quantity $X\in\{r_c,T_c,P_c\}$ is defined with respect to a chosen reference value $\omega_q={\omega_q}_0$ as $\Delta X(\omega_q)=X(\omega_q)-X({\omega_q}_0).
\label{eq:absolute_shift}$ The corresponding fractional shift is expressed as
\begin{equation}
\delta X(\omega_q)
=\frac{\Delta X(\omega_q)}{X({\omega_q}_0)}\times100\%
=\frac{X(\omega_q)-X({\omega_q}_0)}{X({\omega_q}_0)}\times100\%.
\label{eq:fractional_shift}
\end{equation}
The calculated fractional shifts are listed in Table \ref{tab:fractional_shift_percent}. These results provide a quantitative measure of how the critical radius, critical temperature, and critical pressure evolve with the quintessence parameter. The observed trends indicate that variations in $\omega_q$ produce systematic changes in the thermodynamic critical point, highlighting the important role of the quintessence field in determining the phase structure and critical behavior of the black hole system.
\begin{table}[hb]
\centering
\caption{Fractional shifts $\delta X(\omega_q)$ (in percent) relative to $\omega_q=-0.4$, calculated using Eq. (\ref{eq:fractional_shift}).}
\label{tab:fractional_shift_percent}
\begin{tabular}{cccc}
\hline\hline
$\omega_q$ & $|\delta r_c|(\%)$ & $|\delta T_c|(\%)$ & $|\delta P_c(\%)|$ \\
\hline
$-0.5$ & $3.19$ & $6.33$ & $8.04$ \\
$-0.6$ & $7.21$ & $8.21$ & $31.61$ \\
$-0.7$ & $10.68$ & $4.37$ & $74.61$ \\
$-0.8$ & $12.77$ & $5.89$ & $139.38$ \\
$-0.9$ & $12.87$ & $23.00$ & $227.43$ \\
\hline
\end{tabular}
\end{table}

\section*{Acknowledgements}
The work was sponsored by the National Natural Science Foundation of China (Grant Numbers: 12235007, 12375003).

\section*{Data Availability Statement} 
This manuscript has associated data in a data repository. [Authors’ comment: There are no observational data related to this article. The necessary calculations and graphic discussion can be made available on request.]

\section*{Declarations}
Conflict of interest: The authors declare that they have no known competing financial interests or personal relationships that could have appeared to influence the work reported in this paper.

\end{document}